\documentclass[preprint,12pt]{elsarticle}

\usepackage{amssymb,amsmath,natbib,textcomp}

\journal{Journal of Quantitative Spectroscopy \& Radiative Transfer}

\begin{document}

\begin{frontmatter}



\title{A numerical model for multigroup radiation hydrodynamics}


\author{N. Vaytet, E. Audit}

\address{Service d'Astrophysique, CEA/DSM/IRFU/SAp, Centre d'\'{E}tudes de Saclay, L'Orme des Merisiers, 91191 Gif-sur-Yvette, Cedex, France}

\author{B. Dubroca}

\address{CELIA, Universit\'{e} Bordeaux I, CNRS, CEA, 33405 Talence, France}

\author{F. Delahaye}

\address{LERMA, Observatoire de Paris, ENS, UPMC, UCP, CNRS, 5 Place Jules Janssen, 92190 Meudon, France}

\begin{abstract}

We   present  in  this   paper  a   multigroup  model   for  radiation
hydrodynamics  to account  for  variations  of the  gas  opacity as  a
function of frequency. The  entropy closure model ($M_{1}$) is applied
to multigroup radiation transfer in a radiation hydrodynamics code. In
difference from the previous grey  model, we are able to reproduce the
crucial  effects  of  frequency-variable  gas opacities,  a  situation
omnipresent  in physics  and  astrophysics. We  also  account for  the
energy  exchange between  neighbouring  groups which  is important  in
flows with strong velocity divergence. These terms were computed using
a finite volume method in the frequency domain. The radiative transfer
aspect  of   the  method  was  first  tested   separately  for  global
consistency (reversion  to grey model) and against  a well established
kinetic  model through  Marshak  wave tests  with frequency  dependent
opacities.  Very good  agreement  between the  multigroup $M_{1}$  and
kinetic models was  observed in all tests. The  successful coupling of
the  multigroup  radiative  transfer  to the  hydrodynamics  was  then
confirmed through  a second  series of tests.  Finally, the  model was
linked to  a database of opacities for  a Xe gas in  order to simulate
realistic multigroup radiative shocks  in Xe. The differences with the
previous grey models are discussed.

\end{abstract}

\begin{keyword}
Radiative transfer \sep Moment model \sep Multigroup \sep Laboratory astrophysics

\PACS 47.70.Mc

\end{keyword}

\end{frontmatter}


\section{Introduction}\label{sec:intro}

The study of radiative transfer and its interaction with matter has an
extremely wide  range of applications ranging from  medical imagery to
astrophysics. In  many cases, as  for example in  stellar atmospheres,
the radiation is considered as  a physical probe which provides access
to the thermodynamical properties of  the flow through the spectrum of
emission and absorption lines. However, the radiation often has a very
important dynamical role  in the system. It cannot  only be considered
as a passive probe, but as an integral part of the equations governing
the system dynamics.

The equation of radiative transfer (ignoring scattering) is
\begin{equation}\label{eq:trans}
\displaystyle
\left( \frac{1}{c}  \frac{\partial}{\partial t}+ \mathbf{n} \cdot \nabla \right) I(\mathbf{x},t;\mathbf{n},\nu) 
= \sigma_{\nu} \Big( B(\mathbf{x},t,\nu) - I(\mathbf{x},t;\mathbf{n},\nu) \Big)
\end{equation}

where  $I$ is  the  specific  intensity of  the  radiation, $\nu$  the
frequency,   $c$   is  the   speed   of   light,  $\sigma_{\nu}$   the
absorption/emission  coefficient  and  $B$  the  black  body  specific
intensity.  $\mathbf{n}$,  $\mathbf{x}$,  and  $t$  are  the  angular,
spatial  and  temporal   variables,  respectively.  As  the  radiation
intensity depends on seven  variables in three-dimensions, solving the
full  transfer  equation  coupled   to  the  hydrodynamics  to  tackle
radiation hydrodynamics (RHD) problems is still out of reach of modern
computational  architectures, even  with the  remarkable  and constant
increase in computing power.

In order  to overcome this difficulty,  much effort has  been spent in
recent years developing  mathematically less complicated, yet accurate
approximations   to  the   equations  of   radiative   transfer.  Such
approximations         include         diffusion        approximations
\citep{larsen74,larsen02,larsen03}       and       moment       models
\citep{dubroca99,levermore84,muller98,struchtrup98}.   All   of  these
approximations use  frequency and/or angle-integrated  variables which
greatly  simplify  the calculations.  The  approximations  due to  the
angular  integration  have  been   widely  studied  (see  for  example
\citet{olson98}). However, in many situations, the quantities involved
in the  equations of radiative transfer (in  particular the absorption
and scattering coefficients) depend  strongly on frequency, and the so
called `grey'  approximation (integrated  over all frequencies)  is no
longer  appropriate. Only very  recently have  models which  take into
account      variations      in      frequency     been      developed
\citep{turpault05,ramis88,blinnikov00,skartlien00}.     The     common
practise is to split the frequency domain into a finite number of bins
or groups  and the equations  of radiative transfer are  solved within
each group;  this is known  as a multigroup  method. Such a  scheme is
then capable of  allowing for gas opacity variations  in the frequency
domain  providing a  more accurate  description of  radiative transfer
processes. The model  we present in this paper is  an extension of the
moment  model  introduced  in  \citet{turpault05}  which  couples  the
frequency-dependent radiation to the hydrodynamics.

Moment models are obtained  by computing successive angular moments of
the radiative transfer equation.  One obtains a hierarchy of equations
for  moments  of  the  specific intensity.  Basically,  each  equation
describes the evolution of  the n$^{\mathrm{th}}$ moment as a function
of the  divergence of the (n+1)$^{\mathrm{th}}$  moment. For instance,
the equations giving the evolution of the first two moments are
\begin{equation}\label{eq:evolrad}
\begin{array}{lcrcl}
\partial_{t} E_{\nu}          &+&     \nabla \cdot \mathbf{F}_{\nu} &=&   \sigma_{\nu} (4\pi B - c E_{\nu}) \\
\partial_{t} \mathbf{F}_{\nu} &+& c^2 \nabla \cdot \mathbb{P}_{\nu} &=& - \sigma_{\nu} c \mathbf{F}_{\nu} 
\end{array}
\end{equation}
where $E_{\nu}, \mathbf{F}_{\nu}$, and $\mathbb{P}_{\nu}$ are, respectively, the radiative energy density, the radiative energy flux, and the radiative pressure, which are defined in terms of the zeroth, first and second moments of the specific intensity as
\begin{equation}\label{eq:defmoment}
\begin{array}{lcrc}
E_{\nu}          = \displaystyle \frac{1}{c} & \displaystyle \oint &                                  I(\mathbf{x},t;\mathbf{n},\nu) ~ d\Omega & \\
\mathbf{F}_{\nu} =                           & \displaystyle \oint &  \mathbf{n}                    ~ I(\mathbf{x},t;\mathbf{n},\nu) ~ d\Omega & \\
\mathbb{P}_{\nu} = \displaystyle \frac{1}{c} & \displaystyle \oint &  \mathbf{n} \otimes \mathbf{n} ~ I(\mathbf{x},t;\mathbf{n},\nu) ~ d\Omega & .
\end{array}
\end{equation}

The transfer equation is  formally equivalent to an infinite hierarchy
of moment  equations. In order to  have a tractable  moment model, one
must cut  this hierarchy  at some given  order. A closure  relation is
then  needed in  order to  express the  moment of  highest order  as a
function of the others.

In  this paper,  we develop  for the  first time  the coupling  of the
$M_{1}$ moment  model for radiative  transfer to the  hydrodynamics to
create  a multigroup model  for RHD.  A finite  volume method  used to
compute  the   additional  terms   due  to  frequency   variations  is
described. We then present a  series of tests for the multigroup model
studying  first the radiative  transfer alone  and then  the radiation
coupled to the hydrodynamics. The strengths and future developments of
the model are finally discussed.

\section{The multigroup model for radiation hydrodynamics}\label{sec:multigmodel}

\subsection{The monochromatic equations of radiation hydrodynamics}\label{sec:equ_rhd_monoch}

The  equations  of radiation  hydrodynamics  describe  the effects  of
radiative  transfer  on  a   moving  fluid.  The  fluid  evolution  is
determined  by the classical  conservation equations  (mass, momentum,
and  energy) which  are coupled  to the  radiative  transfer equations
(\ref{eq:evolrad})  through source  terms characterizing  the momentum
and energy exchanges between the fluid and the radiation.

In order  to write the RHD equations,  one has to choose  the frame in
which  to  evaluate  the  radiative quantities:  laboratory  frame  or
comoving frame (i.e. the frame  moving with the fluid). The laboratory
frame  is convenient  because the  the  left-hand side  of the  system
remains     hyperbolic      and     thus     globally     conservative
\citep{mihalas01}.  However in  this frame,  interactions  with matter
become complex because of Doppler  and aberration effects that have to
be  incorporated in the  source terms.  On the  other hand,  using the
radiative quantities expressed  in the comoving frame \citep{lowrie01}
adds  non-conservative terms to  the equations, and conversions of the
radiative quantities between comoving and lab frames are required in
order to be compared to observations as any measurement will almost
certainly be carried out in the lab frame.  However the  source terms
coupling matter and radiation remain unaffected by the fluid motions.

We  have  chosen  to  express  radiative quantities  in  the  comoving
frame for the greater simplicity of the source terms. The equations of non-relativistic RHD (to order $u/c$) can then
be written as \citep{lowrie01,mihalas84,buchler79}
\begin{equation}\label{eq:hydro}
\begin{array}{lclcl}
\partial_t \rho             & + & \nabla \cdot (\rho \mathbf{u})                                  & = & 0 \\
\partial_t(\rho \mathbf{u}) & + & \nabla \cdot (\rho \mathbf{u} \otimes \mathbf{u} + p\mathbb{I}) & = &   \displaystyle \int_{0}^{\infty} (\sigma_{\nu}/c) \mathbf{F}_{\nu} d\nu \\
\partial_t e                & + & \nabla \cdot \big(\mathbf{u}(e + p)\big)                        & = & - \displaystyle \int_{0}^{\infty} \Big( \sigma_{\nu} (4\pi B - c E_{\nu}) \\
~                           & ~ & ~                                                               & ~ & - (\sigma_{\nu}/c) \mathbf{u} \cdot \mathbf{F}_{\nu} \Big) ~d\nu
\end{array}
\end{equation}
\begin{equation}\label{eq:ray}
\begin{array}{l}
\partial_{t} E_{\nu}          +       \nabla \cdot \mathbf{F}_{\nu} + \mathbb{P}_{\nu} : \nabla \mathbf{u}     + \nabla \cdot (\mathbf{u} E_{\nu})                  - \partial_{\nu}(\nu \mathbb{P}_{\nu}) : \nabla \mathbf{u} = \sigma_{\nu} (4\pi B - c E_{\nu}) \\
\partial_{t} \mathbf{F}_{\nu} + c^{2} \nabla \cdot \mathbb{P}_{\nu} + \mathbf{F}_{\nu} \cdot \nabla \mathbf{u} + \nabla \cdot (\mathbf{u} \otimes \mathbf{F}_{\nu}) - \partial_{\nu}(\nu \mathbb{Q}_{\nu}) : \nabla \mathbf{u} = - \sigma_{\nu} c \mathbf{F}_{\nu}
\end{array}
\end{equation}
where $\rho$ is the gas density, $u$ the velocity, $e$ the total gas energy, $p$ the gas pressure, and $\mathbb{Q}_{\nu}$ is the third moment of the specific intensity
\begin{equation}\label{eq:defq}
\mathbb{Q}_{\nu} = \oint \mathbf{n} \otimes \mathbf{n} \otimes \mathbf{n} ~ I(\mathbf{x},t;\mathbf{n},\nu) ~ d\Omega ~~~.
\end{equation}
The tensorial contractions are defined by $\mathbb{P} : \nabla \mathbf{u} = \mathbb{P}_{ij} \partial^{i} u^{j}$ and $\mathbb{Q} : \nabla \mathbf{u} = \mathbb{Q}_{ijk} \partial^{i} u^{j}$.

\subsection{The multigroup equations of radiation hydrodynamics}\label{sec:eq_rhd_multi}

Equations  (\ref{eq:hydro}) and (\ref{eq:ray})  are all  Eulerian, but
the radiative quantities are evaluated  in the frame comoving with the
fluid. In a grey model,  system (\ref{eq:ray}) is integrated from 0 to
$\infty$ in  frequency and  the terms involving  frequency derivatives
$\partial_{\nu}(\nu    \mathbb{P}_{\nu})$    and   $\partial_{\nu}(\nu
\mathbb{Q}_{\nu})$ vanish. However, in  a multigroup model these terms
remain  and  are in  fact  of  great  importance; they  govern  energy
transfers between neighbouring groups.

In a multigroup  model, the frequency domain is  divided into a finite
number  of bins  or groups  and the  radiative transfer  equations are
integrated and solved  within each group. The integrals  in the source
terms of the hydrodynamic  equations (\ref{eq:hydro}) then become sums
of   source  terms   over  the   total  number   of   groups.  Systems
(\ref{eq:hydro}) and (\ref{eq:ray}) become
\begin{equation}\label{eq:hydromulti}
\begin{array}{lclcl}
\partial_t \rho             & + & \nabla \cdot (\rho \mathbf{u})                                  & = & 0 \\
\partial_t(\rho \mathbf{u}) & + & \nabla \cdot (\rho \mathbf{u} \otimes \mathbf{u} + p\mathbb{I}) & = &   \displaystyle \sum_{g=1}^{Ng} (\sigma_{Fg}/c) \mathbf{F}_{g} \\
\partial_t e                & + & \nabla \cdot (\mathbf{u}(e + p))                                & = & - \displaystyle \sum_{g=1}^{Ng} \Big( c(\sigma_{Pg} \Theta_{g}(T) - \sigma_{Eg} E_{g}) \\
                            &   &                                                                 &   &  - (\sigma_{Fg}/c) \mathbf{u} \cdot \mathbf{F}_{g} \Big)
\end{array}
\end{equation}
\begin{multline*}
\partial_{t} E_{g} + \nabla \cdot \mathbf{F}_{g} + \nabla \cdot (\mathbf{u} E_{g}) + \mathbb{P}_{g} : \nabla \mathbf{u} 
- \nabla \mathbf{u} : \displaystyle \int_{\nu_{g-1/2}}^{\nu_{g+1/2}}\partial_{\nu}(\nu \mathbb{P}_{\nu}) d\nu \\ 
= c \big( \sigma_{Pg} \Theta_{g}(T) - \sigma_{Eg} E_{g} \big)
\end{multline*}
\begin{multline}\label{eq:raymulti2}
\partial_{t} \mathbf{F}_{g} + c^{2} \nabla \cdot \mathbb{P}_{g} + \nabla \cdot (\mathbf{u} \otimes \mathbf{F}_{g}) 
+ \mathbf{F}_{g} \cdot \nabla \mathbf{u} - \nabla \mathbf{u} : \displaystyle \int_{\nu_{g-1/2}}^{\nu_{g+1/2}}\partial_{\nu}(\nu \mathbb{Q}_{\nu}) d\nu \\ 
= - \sigma_{Fg} c \mathbf{F}_{g}
\end{multline}
with
\begin{equation}\label{eq:groupvar}
X_{g} = \int_{\nu_{g-1/2}}^{\nu_{g+1/2}} X_{\nu} d\nu
\end{equation}
where  $X   =  E$,  $\mathbf{F}$,   $\mathbb{P}$,  $\mathbb{Q}$  which
represent the  radiative energy, flux,  pressure and heat  flux inside
each  group  $g$ which  holds  frequencies  between $\nu_{g-1/2}$  and
$\nu_{g+1/2}$.   $N_{g}$   is  the   total   number   of  groups   and
$\Theta_{g}(T)$  is  the  energy   of  the  photons  having  a  Planck
distribution at  temperature $T$ inside a given  group. The absorption
coefficients  $\sigma_{Pg}$, $\sigma_{Eg}$  and $\sigma_{Fg}$  are the
means of  $\sigma_{\nu}$ inside a  given group weighted by  the Planck
function, the radiative energy and the radiative flux respectively.

In  order  to  integrate  the  previous system,  it  is  necessary  to
introduce a closure relation giving $\mathbb{P}$ and $\mathbb{Q}$ as a
function of $E$ and $\mathbf{F}$.  The closure we have chosen is based
on the $M_1$ model and is presented below.

\subsection{The multigroup $M_{1}$ model}\label{sec:m1model}

The $M_{1}$ model \citep{dubroca99} uses the first two moment equations
(\ref{eq:evolrad}) to approximate the equation of radiative transfer.
It has the great advantage over flux-limited diffusion models 
\citep{minerbo78,levermore81,levermore84} of being valid in both the
diffusion and free-streaming limits while maintaining a directionality
in the propagation of the radiation. Shadows can be created with the $M_{1}$
method while flux-limited diffusion considers the radiative flux to always be colinear
to the radiative temperature gradient, which can result in radiation
propagating around corners \citep{dubroca99,gonzalez07}.

\begin{figure}
\begin{center}
\includegraphics[scale=0.4]{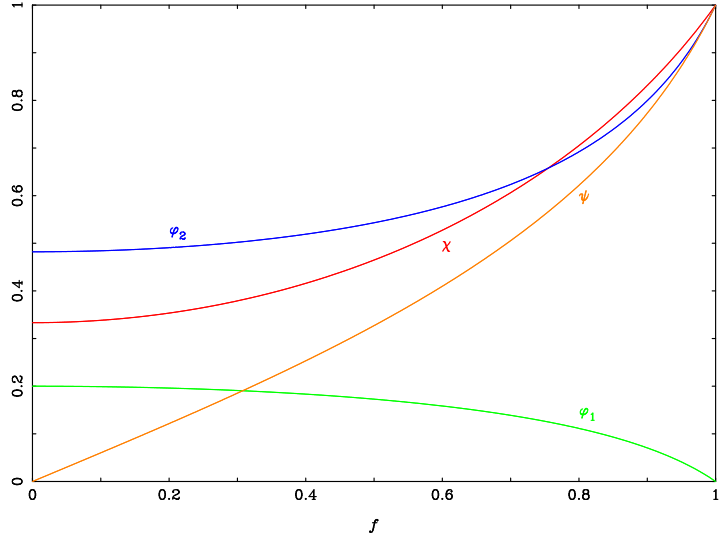}
\caption{$\chi$ (red), $\varphi_{1}$ (green), $\varphi_{2}$ (blue) and $\psi$ (orange) as a function of $f$. $\chi$, $\varphi_{1}$ 
and $\varphi_{2}$ are symmetric with respect to the ordinates axis, $\psi$ is symmetric with respect to the origin.}
\label{fig:chi}
\end{center}
\end{figure}

In the $M_{1}$ model the radiative pressure is expressed as  $\mathbb{P} = \mathbb{D} E$ where
$\mathbb{D}$ is known as the Eddington tensor. The expression for
$\mathbb{D}$ is obtained by minimizing  the  radiative entropy which
yields
\begin{equation}\label{eq:eddtensor}
\mathbb{D} = \frac{1 - \chi}{2} ~ \mathbb{I} + \frac{3\chi - 1}{2} 
~ \frac{\mathbf{F} \otimes \mathbf{F}}{\|\mathbf{F}\|^{2}}
\end{equation}
where
\begin{equation}\label{eq:chi1}
\chi  = \frac{3 + 4 f^{2}}{5 + 2 \sqrt{4 - 3 f^{2}}}
\end{equation}
and $f =  \frac{\| \mathbf{F} \|}{c E}$ is the ratio  of the grey flux
to the flux free-streaming limit,  also known as the reduced flux. The
quantities without the subscript $\nu$ represent quantities integrated
over the  entire frequency range. Note  that by definition  of $E$ and
$\mathbf{F}$, we  have $f \le  1$, which implies that  the radiative
energy is transported  at most at the speed of  light. In one dimension 
we simply have $\mathbb{P} = \chi E$. We have plotted
$\chi$ in Fig.~\ref{fig:chi} as a  function of $f$ (red). This closure
relation recovers the two asymptotic regimes of radiative transfer. In
the free-streaming limit  (i.e. transparent media), we have  $f = 1$
and $\chi = 1$.  On the other  hand, in the diffusion limit, $f = 0$
and   $\chi=1/3$,  which   corresponds  to   an   isotropic  radiation
pressure.

In order to express the radiative heat flux $\mathbb{Q}$ as a function
of  the lower  angular moments  using the  $M_{1}$ closure,  we define
$\mathbb{Q}  = \mathbb{H} E  c$.  Due  to the  symmetry of  the specific intensity
distribution function  around the axis defined by the direction of 
propagation of the radiative flux, it can be shown that
\begin{equation}\label{eq:defHr}
\mathbb{H} = \varphi_{1} ( \mathrm{f}_{i} \delta_{jk} + \mathrm{f}_{j} \delta_{ik} + \mathrm{f}_{k} \delta_{ij} ) + \varphi_{2} ( \mathrm{f}_{i} \mathrm{f}_{j} \mathrm{f}_{k} )
\end{equation}
in which $\mathrm{f}_{i}$ are the components of the reduced flux vector $\mathbf{f} =  \frac{\mathbf{F}}{c E}$,
\begin{multline}\label{eq:phi1}
\varphi_{1} = \frac{(f-2+a)(f+2-a)}{4f(a - 2)^{5}} \Bigg[ 12 \ln \left( \frac{f-2+a}{f+2-a} \right) ( f^{4} + 2af^{2} - 7f^{2} - 4a + 8 )  \\
+ 48f^{3} - 9af^{3} - 80f + 40af \Bigg]
\end{multline}
and
\begin{multline}\label{eq:phi2}
\varphi_{2} = \frac{1}{f^{3}(a - 2)^{5}} \Bigg[ 60 \ln \left( \frac{f-2+a}{f+2-a} \right) ( - f^{6} + 15f^{4} - 3af^{4} + 15af^{2} - 42f^{2} \\
- 16a + 32 ) + 54af^{5} - 465f^{5} - 674af^{3}  + 2140f^{3} + 1056af - 2112f \Bigg]
\end{multline}
where $a  = \sqrt{4-3f^{2}}$. We plot  $\varphi_{1}$ and $\varphi_{2}$
as a function of $f$ in Fig.~\ref{fig:chi} (green and blue). Note that
in one  dimension, $\mathbf{f}  = (f,0,0)$ and  we have  $\mathbb{Q} =
\psi E c$ where
\begin{equation}\label{eq:psi1d}
\psi = 3 \varphi_{1} f + \varphi_{2} f^{3}
\end{equation}
and  $\psi$ is  plotted  in Fig.~\ref{fig:chi}  (orange).  This  fully
defines  the evolution  of the  radiative energy  and pressure  of the
model coupled to the hydrodynamics of the system.

A natural way to extend this closure to a multigroup model would be to
minimize the total radiative entropy, which is a rather complex procedure.
However, \citet{turpault05} has shown  that applying  inside each  group a
closure  formally equivalent  to  the $M_{1}$  closure  leads to  almost
indistinguishable  results; a strategy which we have therefore adopted
for its greater simplicity.  We define  for each  group the radiative
pressure as $\mathbb{P}_{g} = \mathbb{D}_{g} E_{g}$ where
\begin{equation}\label{eq:eddtensorg}
\mathbb{D}_{g} = \frac{1 - \chi_{g}}{2} ~ \mathbb{I} + \frac{3\chi_{g} - 1}{2} ~ \frac{\mathbf{F}_{g} \otimes \mathbf{F}_{g}}{\|\mathbf{F}_{g}\|^{2}} ~~~,
\end{equation}
\begin{equation}\label{eq:chi1g}
\chi_{g}  = \frac{3 + 4 f_{g}^{2}}{5 + 2 \sqrt{4 - 3 f_{g}^{2}}}
\end{equation}
and $f_{g}  = \frac{\|  \mathbf{F}_{g} \| }{c  E_{g}}$. The  heat flux
$\mathbb{Q}_{g}  = \mathbb{H}_{g}  E_{g} c$  is computed  in  the same
manner.

\subsection{A finite volume method for the frequency derivatives}\label{sec:finVol}

The only terms which were not included in our previous grey RHD models
\citep{gonzalez07} are the terms in (\ref{eq:raymulti2}) involving the
frequency   differentials. In order to evaluate these terms, we adopt
a finite volume method in the frequency dimension. We present here this
method in the one-dimensional case, but its extension  to several
dimensions  is trivial.  Retaining  only the time  and  frequency
derivatives  of  the  radiative  energy and  flux equations of system
(\ref{eq:ray}), we obtain
\begin{equation}\label{eq:dnu}
\begin{array}{rcc}
\partial_{t} E_{\nu}           - \mathcal{D} ~ \partial_{\nu} ( \nu \mathbb{P}_{\nu} ) &=& 0 \\
\partial_{t}  \mathbf{F}_{\nu} - \mathcal{D} ~ \partial_{\nu} ( \nu \mathbb{Q}_{\nu} ) &=& 0
\end{array}
\end{equation}
where $\mathcal{D} = \nabla \cdot  \mathbf{u}$. We now assume that the
frequency  group boundaries  ($\nu_{g\pm1/2}$) are  equivalent  to the
volume elements'  boundaries in the frequency  dimension. The  finite
volume  discretization  of (\ref{eq:dnu}) gives 
\begin{equation}\label{eq:findiff2}
\begin{array}{rcc}
\displaystyle \frac{E_{g}^{n+1} - E_{g}^{n}}                  {\Delta t} - \mathcal{D} ~ \Big( \nu_{g+1/2} \mathbb{P}_{g+1/2}^{n} - \nu_{g-1/2} \mathbb{P}_{g-1/2}^{n} \Big) &=& 0 \\
\displaystyle \frac{\mathbf{F}_{g}^{n+1} - \mathbf{F}_{g}^{n}}{\Delta t} - \mathcal{D} ~ \Big( \nu_{g+1/2} \mathbb{Q}_{g+1/2}^{n} - \nu_{g-1/2} \mathbb{Q}_{g-1/2}^{n} \Big) &=& 0
\end{array}
\end{equation}
where   $\mathbb{P}_{g\pm1/2}$  and  $\mathbb{Q}_{g\pm1/2}$   are  the
radiative   pressures   and  heat   flux   evaluated   at  the   group
interfaces. The Jacobian matrix $\mathcal{J} $ of the hyperbolic system
(\ref{eq:dnu}) is given by
\begin{equation}
\mathcal{J} = -\mathcal{D}\nu \tilde{\mathcal{J}} ~~~ \text{with} ~~~
\tilde{\mathcal{J}} = \frac{\partial\left( \begin{array}{c}  \mathbb{P}_{\nu} \\ \mathbb{Q}_{\nu} \end{array}  \right) }
                   {\partial\left( \begin{array}{c}  E_{\nu}          \\ \mathbf{F}_{\nu} \end{array}  \right) }
            = \left(  \begin{array}{cc} \chi -f\chi' & \displaystyle \frac{\chi'}{c} \\ c\psi - cf\psi' & \psi' \end{array} \right)
\end{equation}
where $'$  denotes derivatives with respect to $f$. It can be shown that
the  trace and  the  determinant  of $\tilde{\mathcal{J}}$  are both strictly
positive. The eigenvalues of system (\ref{eq:dnu}) are thus always of the same sign (i.e. opposite to that of $\mathcal{D}$),
which enables us to use a standard upwind scheme with respect to $\mathcal{D}$
to calculate the values for $\mathbb{P}$ and $\mathbb{Q}$ at the group
interfaces. This yields
\begin{equation}\label{eq:PQval}
\displaystyle
\begin{array}{ll}
\mathbb{X}_{g-1/2} & = ~\left\{ \begin{array}{ll} \mathbb{X}_{g  }/\Delta \nu_{g  } & \text{if} ~ \mathcal{D} > 0 \\ \mathbb{X}_{g-1}/\Delta \nu_{g-1} & \text{if} ~ \mathcal{D} \leq 0 \end{array} \right. \\~\\
\mathbb{X}_{g+1/2} & = ~\left\{ \begin{array}{ll} \mathbb{X}_{g+1}/\Delta \nu_{g+1} & \text{if} ~ \mathcal{D} > 0 \\ \mathbb{X}_{g  }/\Delta \nu_{g  } & \text{if} ~ \mathcal{D} \leq 0 \end{array} \right.
\end{array}
\end{equation}
where $\mathbb{X} = \mathbb{P}$ or $ \mathbb{Q}$. This shows  that the
radiative energy  and flux are  advected from one
group to the  other depending on the sign  of the velocity divergence.
It  is straightforward  to  show  that the  inclusion  of these  terms
preserves the flux limitation condition $|f| \le 1$ as long as $\psi <
\chi$,  which is always  true (see  Fig.~\ref{fig:chi}).  It  would of
course  be  possible to  use  higher  order  schemes to  evaluate  the
quantities at the group interfaces by computing slopes using the usual
methods.   For the  sake of  conciseness this  will not  be explicited
here.

\section{Numerical method and tests }\label{sec:results}

\subsection{Numerical method}

In this section,  we briefly present our global  strategy to integrate
the coupled RHD system (\ref{eq:hydromulti})-(\ref{eq:raymulti2}) (the
method is identical to the one reported in \citet{gonzalez07} apart from
the terms involving the frequency differentials which were not included).
In order to have a tractable time step, the radiative transport needs to
be treated implicitly. However, it  is most of the time more efficient
to retain an explicit scheme for the hydrodynamics. We therefore use the
following   splitting  scheme.

In the first step   the  hydrodynamics   system
(\ref{eq:hydromulti}) is  solved explicitly without  the source terms.
It is integrated using  a classical second order MUSCL-Hancock scheme.
In the second  step, the  radiation and  the coupling  terms are solved
implictly.   The  terms  in  system (\ref{eq:raymulti2}) involving  
frequency  derivatives   are discretized as  presented above and using the
velocity divergence from the  hydrodynamic solver. The velocity coming
from  the hydrodynamic  solver is  also used  to discretize  the other
terms  involving velocity  derivatives. For  the  hyperbolic radiative
term (two  first left hand-side terms  of system (\ref{eq:raymulti2}))
we use a HLLC solver with an asymptotic preserving correction in order
to   recover   properly   the   diffusion  limit \citep{berthon07}. System
(\ref{eq:raymulti2})  is solved  implicitly with  the source  terms of
system (\ref{eq:hydromulti}) using a Raphson-Newton procedure.
\begin{align}\label{eq:hydromultinumerical}
\text{Step~1} & \left\{
\begin{array}{lclcl}
\partial_t \rho             & + & \nabla \cdot (\rho \mathbf{u})^{n}                                  & = & 0 \\
\partial_t(\rho \mathbf{u}) & + & \nabla \cdot (\rho \mathbf{u} \otimes \mathbf{u} + p\mathbb{I})^{n} & = & 0 \\
\partial_t e                & + & \nabla \cdot (\mathbf{u}(e + p))^{n}                                & = & 0
\end{array}
\right. \\
~ & ~ \nonumber \\
\text{Step~2} & \left\{
\begin{array}{l}
\partial_{t} E_{g}           +  \nabla \cdot \mathbf{F}^{n+1}_{g}        +  \nabla \cdot (\mathbf{u} E_{g})^{n+1}                   +  (\mathbb{P}_{g} : \nabla \mathbf{u})^{n+1}     \\
~~~~~~~~~~~~~~~ -  (\nabla \mathbf{u} : \tilde{\mathbb{P}}_{g})^{n+1}  =  c \big( \sigma_{Pg} \Theta_{g}(T)               -  \sigma_{Eg} E_{g} \big)^{n+1} \\
\partial_{t} \mathbf{F}_{g}  +  c^{2} \nabla \cdot \mathbb{P}^{n+1}_{g}  +  \nabla \cdot (\mathbf{u} \otimes \mathbf{F}_{g})^{n+1}  +  (\mathbf{F}_{g} \cdot \nabla \mathbf{u})^{n+1}  \\
~~~~~~~~~~~~~~~ -  (\nabla \mathbf{u} : \tilde{\mathbb{Q}}_{g})^{n+1}  =  - (\sigma_{Fg} c \mathbf{F}_{g})^{n+1}          ~  ~ \\
\partial_{t} e               =  - \displaystyle \sum_{g=1}^{Ng}          \Big(  c (\sigma_{Pg} \Theta_{g}(T)  -  \sigma_{Eg} E_{g})   -  (\sigma_{Fg}/c) \mathbf{u} \cdot \mathbf{F}_{g} \Big)^{n+1} \\
\partial_{t} (\rho u)        =    \displaystyle \sum_{g=1}^{Ng}          \Big(  (  \sigma_{Fg}/c) \mathbf{F}_{g} \Big)^{n+1}     
\end{array}
\right.
\end{align}
where
\begin{equation}
\tilde{\mathbb{P}}_{g} = \int_{\nu_{g-1/2}}^{\nu_{g+1/2}}\partial_{\nu}(\nu \mathbb{P}_{\nu}) d\nu ~~~\text{and}~~~ \tilde{\mathbb{Q}}_{g} = \int_{\nu_{g-1/2}}^{\nu_{g+1/2}}\partial_{\nu}(\nu \mathbb{Q}_{\nu}) d\nu ~~~.
\end{equation}

The discretization of the divergence of a quantity $\mathcal{U}$
in cell $i$ is done following
\begin{equation}\label{eq:discretizedflux}
\nabla \cdot \mathcal{U} = \frac{\mathcal{U}^{*}_{i+\frac{1}{2}} - \mathcal{U}^{*}_{i-\frac{1}{2}}}{\Delta x}
\end{equation}
where the state $\mathcal{U}^{*}_{i+\frac{1}{2}}$ is the value of the quantity $\mathcal{U}$
(hydrodynamic or radiative)
at the interface $i+\frac{1}{2}$ (between cell $i$ and $i+1$) solution to the Riemann problem
with left and right states $\mathcal{U}^{-}_{i+\frac{1}{2}}$ and $\mathcal{U}^{+}_{i+\frac{1}{2}}$, respectively.
For a first order scheme $\mathcal{U}^{-}_{i+\frac{1}{2}} = \mathcal{U}_{i}$
and $\mathcal{U}^{+}_{i+\frac{1}{2}} = \mathcal{U}_{i+1}$. For a second order
scheme, the values of $\mathcal{U}$ are lineraly extrapolated to the interfaces
using local gradients.

It     is     of      course     possible     to     solve     systems
(\ref{eq:hydromulti})-(\ref{eq:raymulti2})  without splitting  using a
fully implicity  scheme. We have tried  this in one  dimension and did
not  find  any  significant  differences  with  the  splitting  scheme
presented above.

Calculating  $\varphi_{1}$ and  $\varphi_{2}$ for  every grid  cell at
every  timestep is computationally  demanding due  to the  presence of
logarithm  and  power  functions,  and  we  have  thus  tabulated  the
functions using 100  points which are read in once by  the code at the
beginning of a run. A specific value of $\varphi_{1}$ or $\varphi_{2}$
is then  found using a  Hermitian cubic spline interpolation  which is
very fast  and accurate; the  errors between the interpolated  and the
true values are less than 0.01\% throughout.

In  this section we  validate the  method for  multigroup RHD  using a
series of tests.  For this  purpose the numerical scheme skeched above
have been implemented in a one-dimensional Lagrangian hydrodynamic
code.

The boundary conditions are implemented using two ghost cells at the edges of the
grid. These  ghost cells are filled using various physical constraints
such as null gradient, reflexive boundary or user  imposed conditions.

We use  a step by  step progression in  our test sequence in  order to
verify   each  aspect   of  the   method  with   increasingly  complex
problems. We  first make sure  that the multigroup transfer  model (no
hydrodynamics included) is equivalent to a grey model if the opacities
are independent  of frequency. We  then test the multigroup  aspect of
the method with frequency dependent opacities. We compare the results
of these  tests to a  well-established kinetic model which  solves the
equation    of    radiative    transfer   (\ref{eq:trans})    directly
\citep{charrier03}.   Thirdly,  we  investigate  the coupling  of  the
radiative    transfer    to    the    gas   motion    using    `frozen
hydrodynamics'. Finally, we perform full RHD tests.

\subsection{Marshak waves}\label{sec:marshak}

\subsubsection{Classical grey Marshak wave}\label{sec:marshG}

\begin{figure}[!ht]
\centering
\includegraphics[scale=0.5]{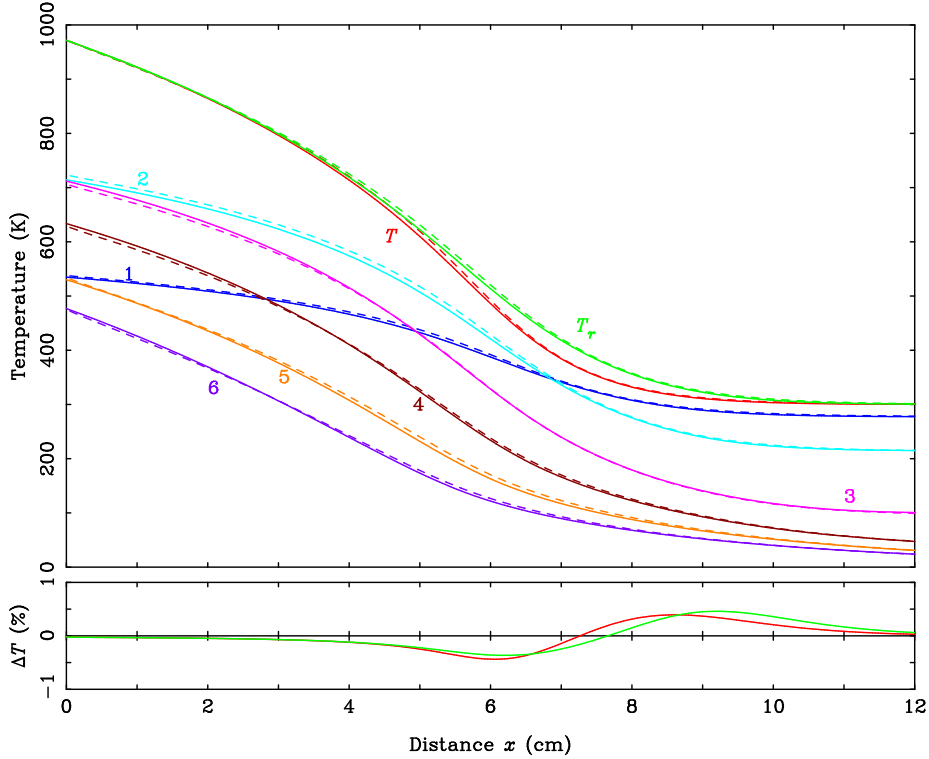}
\caption{Top panel: Gas and radiative temperatures in the grey Marshak
wave test for $\kappa(\nu) = 1000~\text{cm}^{2}~\text{g}^{-1}$ at time
$t = 1.36 \times 10^{-7}$ s.  The solid curves are from the mutligroup
$M_{1}$ model and  the dashed curves represent the  kinetic model. The
red curve marked $T$ is the gas temperature and the green curve marked
$T_{r}$  is   the  total   radiative  temperature  (summed   over  all
groups).  The  other coloured  curves  marked  1  to 6  represent  the
radiative  temperatures inside  each group.  Bottom  panel: percentage
difference between the $M_{1}$ grey  and multigroup models for the gas
temperature (red) and the radiative temperature (green).}
\label{fig:marshG}
\end{figure}

Our first test is to check that the multigroup model reduces to a grey
$M_{1}$ model for a gas with a frequency-independent opacity. We run a
Marshak wave simulation, where the gas inside the grid is at rest with
a uniform density $\rho  = 10^{-3}~\text{g~cm}^{-3}$, temperature $T =
300$  K  in equilibrium  with  the  radiation  and opacity  $\kappa  =
1000~\text{cm}^{2}~\text{g}^{-1}$  independent  of  frequency,  noting
that $\sigma = \kappa \rho$. The  specific heat capacity of the gas is
set          so          that          $\rho          C_{V}          =
10^{-3}~\text{erg~g~cm}^{-3}~\text{K}^{-1}$.  The planar  grid extends
from 0 to  20 cm, using 500 cells.  Boundary conditions: the radiative
energy inside the  left and right ghost cells is that  of a black body
at 1000 K and 300 K,  respectively. The radiative flux inside the left
and right ghost cells is zero. We ran two simulations; the first using
a  single frequency  group from  0 to  $\infty$ (grey  model)  and the
second using  five frequency groups evenly  spaced between $\nu  = 0 -
1.5\times10^{14}~\text{s}^{-1}$ plus a sixth  group to cover the range
$1.5\times10^{14}~\text{s}^{-1}$ to $\infty$. The results are shown in
Fig.~\ref{fig:marshG}  (solid lines),  compared to  the  kinetic model
(dashed lines), at a time $t = 1.36 \times 10^{-7}$ s.

The radiative temperature inside a particular group is defined by
\begin{equation}
T_{r}^{g} = \left( \frac{E_{g}}{a_{R}} \right)^{1/4}
\end{equation}
and the total radiative temperature is
\begin{equation}
T_{r} = \left( \sum_{g=1}^{Ng} E_{g} / a_{R} \right)^{1/4} ~~~.
\end{equation}
In the top  panel, the curves from the  multigroup simulation (and the
kinetic  model)  are plotted.  The  curves  representing  the gas  and
radiative temperatures  for the mono- and  multigroup simulations were
virtually  indistinguishable  and we  show  the percentage  difference
between them  in the  bottom panel. Note  that the  differences remain
below  0.5\%  throughout.  This   shows  that  the  multigroup  scheme
consistently  reduces  to  a  grey  model  for  frequency  independent
opacities.

The kinetic  model solves the  equation of transfer directly  using in
this case 100 spatial zones,  64 directions and 64 frequency bins (all
the results  from the  kinetic model have  been tested  for resolution
convergence).  For a  moment  model, the  total radiative  temperature
(summed over all  groups; bright green) and the  gas temperature (red)
are   in  excellent   agreement  with   their   kinetic  counterparts,
illustrating the validity of  the $M_{1}$ model for radiative transfer
and proving that  the multigroup model consistently reverts  to a grey
model in the case of frequency-independent opacities.

\subsubsection{Multigroup Marshak wave with frequency dependent opacities}\label{sec:marshM}

\begin{figure}[!ht]
\centering
\includegraphics[scale=0.5]{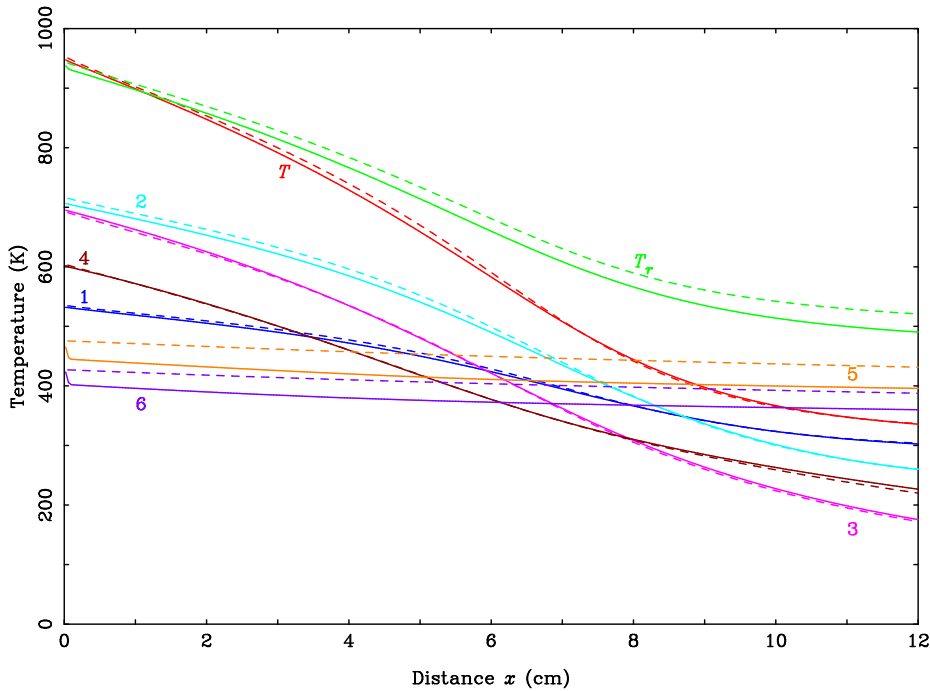}
\caption{Same  as  in  Fig.~\ref{fig:marshG}  but  in the  case  of  a
frequency dependent opacity.}
\label{fig:marshM}
\end{figure}

As a second step, we consider a frequency variable opacity in order to
assess its effect on the Marshak  wave test. The setup is identical to
the grey test  above, but the opacities  in the groups 1 to  6 are (in
$\text{cm}^{2}~\text{g}^{-1}$)  1000,   750,  500,  250,   10  and  10
respectively.  We also used  50 extra  cells with  steadily increasing
widths  at the  right end  of the  grid in  order to  ensure  that the
radiation in  the low-opacity groups does  not have time  to reach the
right edge of the grid. The first 500 zones are the same as above, but
the  total  grid  size  is  $\sim  9$ m.  The  results  are  shown  in
Fig.~\ref{fig:marshM} (solid lines).
The gas and radiation temperatures  $T$ and $T_{r}$ are different from
the ones  in the  first test.  The radiation in  the groups  with weak
opacities (notably groups 5 and 6) has crossed the entire grid and has
heated the gas at the right edge (the gas temperature at that point is
now 330  K). The radiation  in the groups  3 and 4 has  also travelled
further than in the previous test but not as far as groups 5 and 6. We
note that  the radiative temperature of  group 1 at the  right edge is
slightly higher than in the previous test (just above 300 K as opposed
to 275 K). Since its opacity is unchanged, this shows that the gas has
been heated by  the radiation in the other  groups and has re-radiated
some of  its energy into  group 1. The  curves from the  kinetic model
(using 400 cells, 100 directions and 512 frequencies) are also plotted
(dashed  lines). There  is  an extremely  good  agreement between  the
multigroup   and  kinetic  gas   temperatures.  The   total  radiative
temperatures differ somewhat  more than in the previous  test and this
difference is due to larger  discrepancies in the low opacity groups 5
and 6.  $T_{r}^{5}$ and  $T_{r}^{6}$ are very  close to  their kinetic
counterparts at  the left edge of  the grid but then  drop rapidly and
stabilize  to a  lower  value.  This is  a  boundary condition  effect
explained by the fact that when differences between the left and right
fluxes are large (which is the case at the domain boundaries since the
flux in the ghost cells is set to zero)  the $M_{1}$ model becomes less accurate. A
solution to this issue would be to consider an additional third moment
equation or to  solve two half equations (one  for the flux travelling
towards the  left and the other  towards the right)  for the radiative
flux \citep{dubroca02}.

\subsubsection{Multigroup Marshak wave with frequency and temperature dependent opacities}\label{sec:marshMT}

\begin{figure}[!ht]
\centering
\includegraphics[scale=0.5]{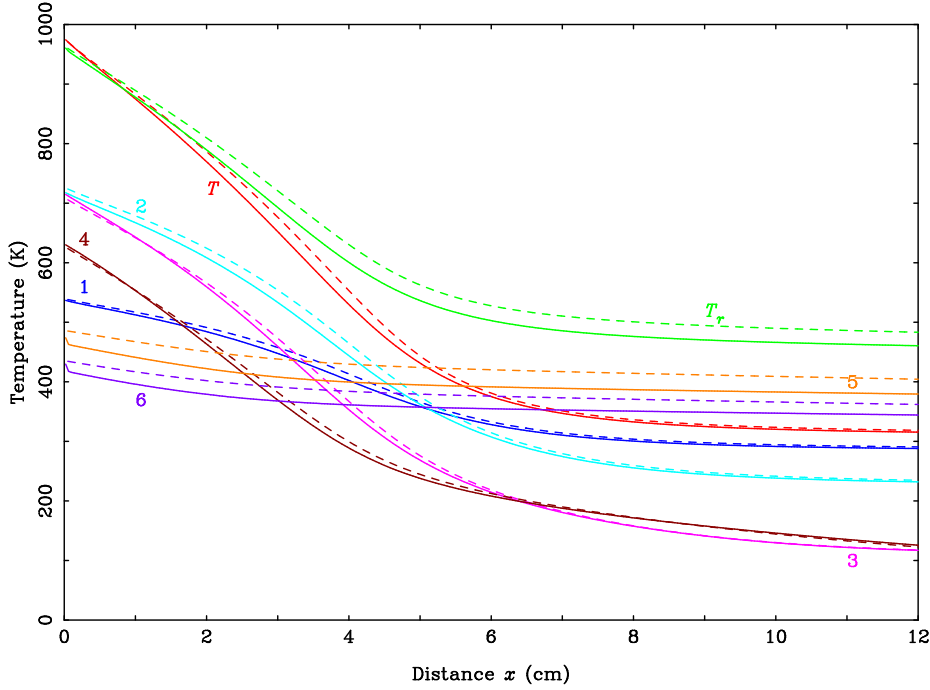}
\caption{Same  as   in  Fig.~\ref{fig:marshM}  but  in   the  case  of
$\kappa(\nu) = \kappa_{0}(\nu) ( T / T_{0})^{3/2}$.}
\label{fig:marshMT}
\end{figure}

In our third test we  use frequency variable opacities which also vary
with temperature. The setup is identical to the multigroup test above,
but the opacities are set to
\begin{equation}
\kappa_{g} = \kappa_{0g} \left( \frac{T}{T_{0}} \right)^{3/2}
\end{equation}
where $T_{0} = 300$  K and $\kappa_{0g}$ in the groups 1  to 6 are the
same     as      in     the     previous      test,     namely     (in
$\text{cm}^{2}~\text{g}^{-1}$)  1000,  750,   500,  250,  10  and  10,
respectively. The  grid setup is  identical to the previous  test. The
results are shown in Fig.~\ref{fig:marshMT}.
This  time, as  the gas  temperature $T$  increases, the  opacity also
increases and the radiation is  absorbed much more rapidly, the effect
being  the most  noticeable  for groups  1  to 4  where opacities  are
high. We see once again  an excellent agreement between the multigroup
$M_{1}$  model  and the  kinetic  model  which  used 512  frequencies,
especially for the gas temperature.

\subsection{Radiation traversing a region with strong velocity variations}\label{sec:usinu}

\begin{figure}[!ht]
\centering
\includegraphics[scale=0.5]{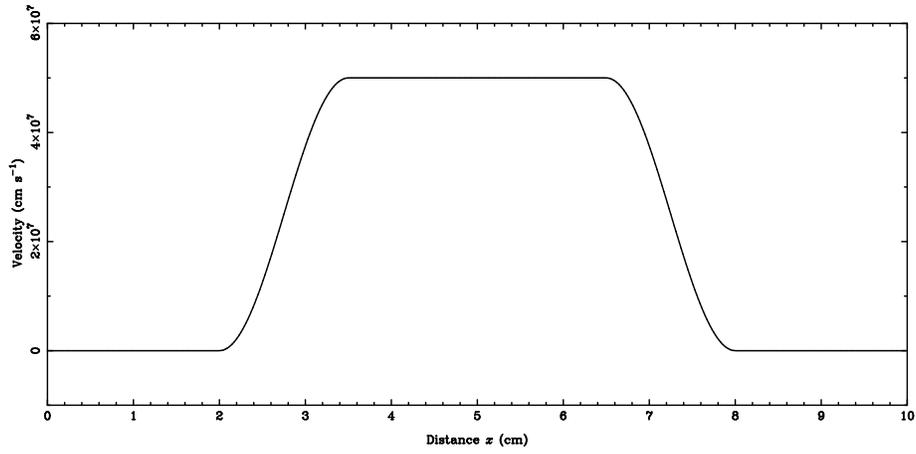}
\caption{Gas velocity as a function of $x$.}
\label{fig:usinu}
\end{figure}

The aim of  this next test is to study  energy exchange between groups
due to Doppler effects when  strong velocity variations are present in
the  fluid.   We  perform  this  test  in vacuum  ($\rho  =  \kappa  =
0$). Radation is cast from  the left side into the computation volume,
with a  black-body spectrum  at $T_{r}  = 1000$ K  and a  unit reduced
flux. The size of  the box is $L = 10$ cm  for 50 cells.  The velocity
is set to obey the following law
\begin{equation}
u(x) ~=~ \left\{\begin{array}{rll}
  0 & ~                                                                & \text{if} ~~ x < x_{0}            \\
  A & \sin^{2} \left( \displaystyle \frac{2\pi}{l} (x - x_{0}) \right) & \text{if} ~~ x_{0} \leq x < x_{1} \\
  A & ~                                                                & \text{if} ~~ x_{1} \leq x < x_{2} \\
  A & \sin^{2} \left( \displaystyle \frac{2\pi}{l} (x - x_{0}) \right) & \text{if} ~~ x_{2} \leq x < x_{3} \\
  0 & ~                                                                & \text{if} ~~ x > x_{3}
\end{array} \right.
\end{equation}
where  $A =  5\times  10^{7} ~\text{cm}~\text{s}^{-1}$,  $l  = 6$  cm,
$x_{0} =  2$ cm, $x_{1} =  3.5$ cm, $x_{2} =  6.5$ cm, $x_{3}  = 8$ cm
(see Fig.~\ref{fig:usinu}). We used 20 equally spaced frequency groups
in the range $0 \rightarrow 2 \times 10^{14}$ Hz, plus a last group to
hold frequencies  in the range $2 \times  10^{14} \rightarrow \infty$.
The radiative temperature  at the boundaries is kept  constant at 1000
K,  and the  radiative reduced  flux is  maintained at  $f =  1$.  The
system is left to evolve until stationarity is reached.

\begin{figure}[!ht]
\centering
\includegraphics[scale=0.5]{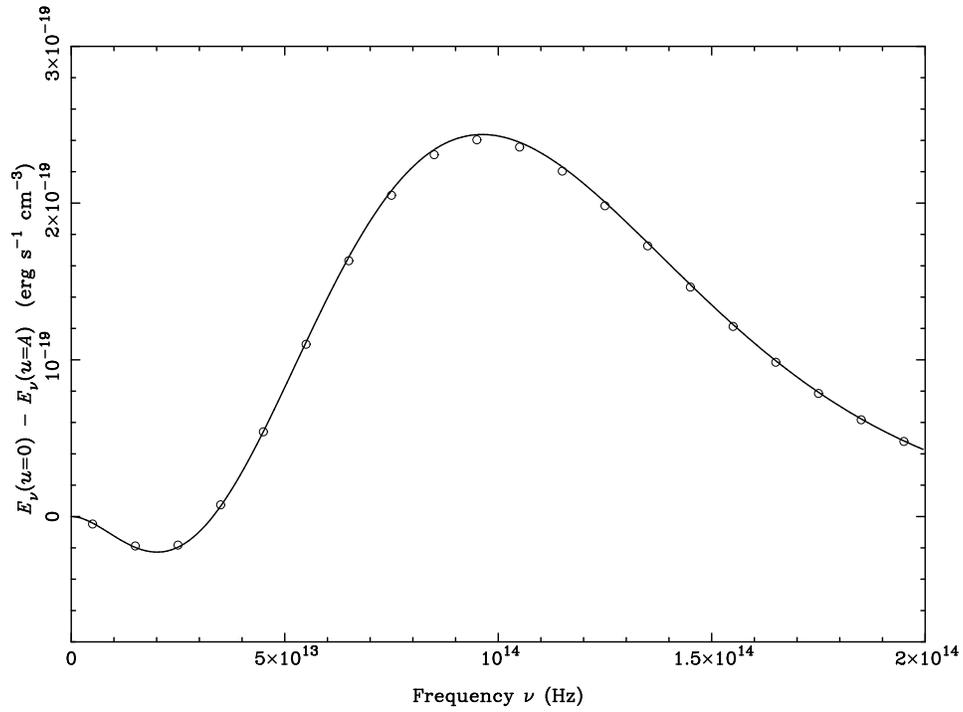}
\caption{Difference in radiative energies between a stationary ($u = 0$) and a moving ($u = A$) black body as a function of frequency. The solid line is the analytical solution, the circles are the numerical solution.}
\label{fig:doppler}
\end{figure}

The difference in radiative energies $E_{\nu}$ between the fixed ($u =
0$)   and   the    moving   ($u   =   A$)   regions    is   shown   in
Fig.~\ref{fig:doppler}. The circles are  the group average numerical solution. 
The solid line  is the  analytical solution,
which  is obtained by  applying a  Doppler shift  in frequency  to the
spectrum
\begin{equation}
\nu' = \gamma \nu \left(1 - \frac{u}{c} \right)
\end{equation}
where
\begin{equation}
\gamma = \frac{1}{\sqrt{1 - \left( \displaystyle \frac{u}{c} \right)^2}} ~~~.
\end{equation}
It is clearly visible in  Fig.~\ref{fig:doppler} that due to the  frequency shift of
the black body spectrum, the  first three frequency groups have gained
energy while  the remaining groups  have lost energy.  The discrepency
between the  analytical and  numerical solution (both  averaged within
frequency groups) is  of the order of one  percent throughout. We have
also performed the test with 10 and 40 groups which did not change the
errors significantly.

\subsection{Velocity gradient with frequency dependent opacities}\label{sec:ugrad}

\begin{figure}[!ht]
\centering
\includegraphics[scale=0.5]{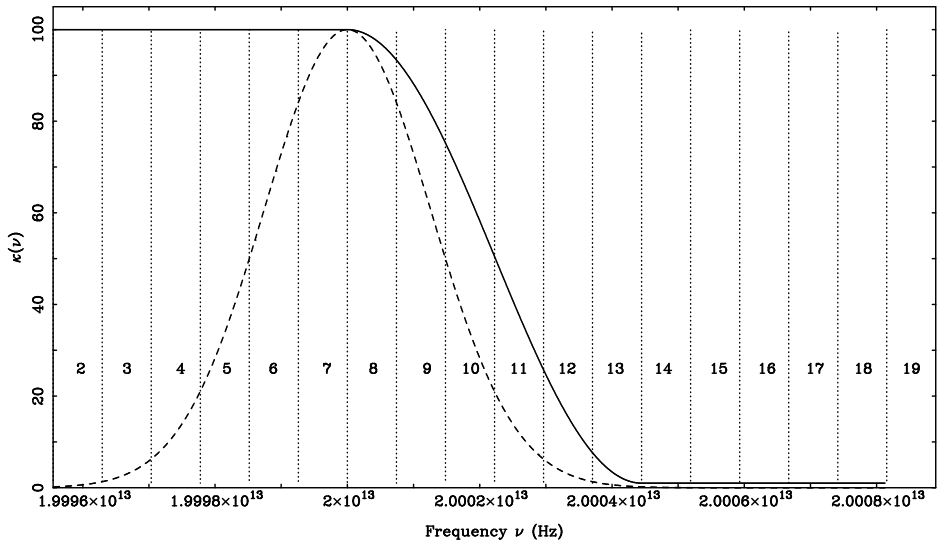}
\caption{Gas    opacity    as    a    function   of    frequency    in
$\mathrm{cm}^{2}~\mathrm{g}^{-1}$  (solid   line).  Intensity  of  the
injected  radiation (normalised;  dashed).  The FWHM  of the  Gaussian
radiative intensity profile measures 2/3 of the width of the opacities
transition  region.  The vertical  black  dotted  lines represent  the
frequency groups identified by their numbers.}
\label{fig:fgroups}
\end{figure}

In this test, we are interested in studying the ability of the code to
handle frequency-variable opacities and  Doppler shifts in a flow with
strong velocity gradients. The size of the box is 1.0 cm for 100 cells
and is initially  filled with a gas  at $T = 3$ K  in equilibrium with
radiation and a velocity $u =  \mathcal{D} x$. As a first step, we set
$\mathcal{D} = 0$. In this  test, the hydrodynamics are frozen and the
gas density is set to $\rho = 1/(\mathcal{C}x)~\text{g~cm}^{-3}$ where
$\mathcal{C}  =  10^{7}$ (see  below).  The  gas  opacity varies  with
frequency: $\kappa(\nu) =  100~\text{cm}^{2}~\text{g}^{-1}$ for $\nu <
2\times10^{13}~\text{s}^{-1}$         and        $\kappa(\nu)        =
1~\text{cm}^{2}~\text{g}^{-1}$           for           $\nu          >
2\times10^{13}~\text{s}^{-1}$,  with a  smooth transition  between the
two regimes of width $\Delta \nu = 4.5\times10^{9}~\text{s}^{-1}$ (see
Fig.~\ref{fig:fgroups}).
We use  20 frequency  groups to sample  the opacities. We  then inject
from the left hand side  a radiation with a Gaussian intensity profile
with a  FWHM measuring  2/3 of the  width of the  opacities transition
region  which  comprised  the same  energy  as  a  1000 K  black  body
radiation.

\begin{figure}[!ht]
\centering
\includegraphics[scale=0.7]{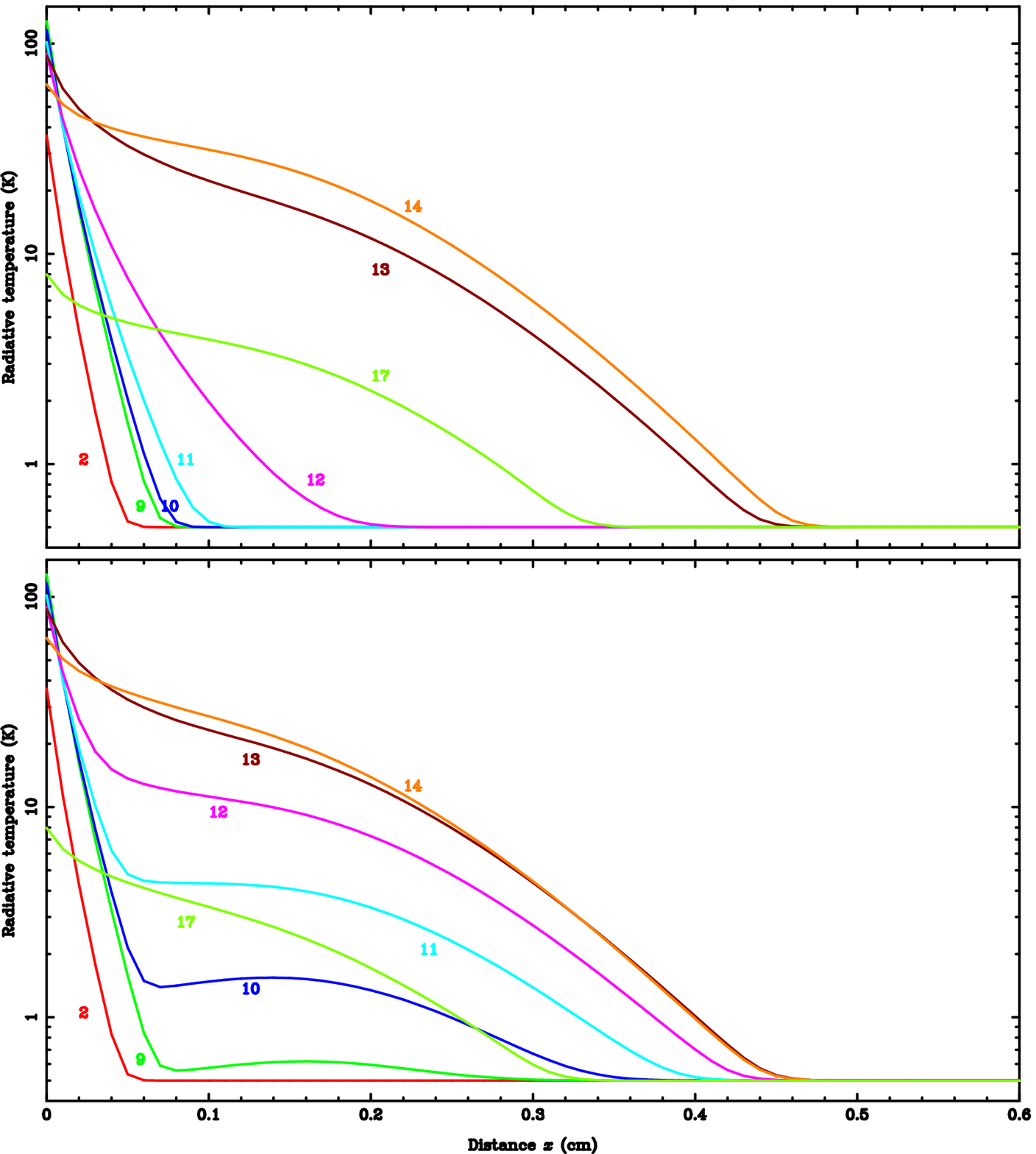}
\caption{Group radiative temperature at time $t = 4.8\times10^{-12}$ s
in the gradient test for a  null velocity (top) and for a gas velocity
which increases linearly with distance (bottom).}
\label{fig:ugrad}
\end{figure}

The  radiative temperatures inside  the separate  groups are  shown in
Fig.~\ref{fig:ugrad}  (top),  where   only  the  relevant  groups  are
presented.
The radiation  in the first 11  groups is rapidly absorbed  by the gas
which has  a high  opacity at these  frequencies, while in  the higher
frequency   groups,   the    radiation   propagates   rapidly   in   a
quasi-transparent medium.  The presence of radiative  energy in groups
13  and above  shows that  the  gas has  been heated  by the  incoming
radiation and has re-radiated some of its energy. Since the heated gas
radiates as a black body, the radiation fills all the groups which are
very narrow  compared to  the width of  a Planck  curve at $T  = 1000$
K. As the opacity is weak  in the high groups, the radiation there can
propagate freely towards the right edge of the box. The lower groups 1
to 3 are  also filled by the black body  radiation but their radiation
cannot escape due to the strong opacities.

In order to  study the effects of velocity  gradients on the radiation
transport, we ran a simulation with the velocity gradient $\mathcal{D}
= 10^{7}~\text{s}^{-1}$. In this case, $\mathcal{D} = \mathcal{C}$ and
a permanent  regime is achieved.  The hydrodynamics are  still frozen,
which is  justified by the fact  that even the fastest  gas would only
have time to move a very small distance ($5\times10^{-5}$ cm) compared
to the box size (1 cm) over the simulation time of $5 \times 10^{-12}$
s.  The results are  shown in  Fig.~\ref{fig:ugrad} (bottom).  We note
that this time,  only the radiation in the first  8 groups is absorbed
and the radiation in groups  9 and above propagates freely. This shows
that the radiation  in the intermediate groups 9,  10 and 11 (covering
the    opacities   transition    region)   was    initially   slightly
absorbed. Then, as  the Doppler frequency shift increases  (due to the
increasing  velocity),  the  radiation   moves  to  groups  of  higher
frequencies where the opacity is  much lower and the radiation is thus
able to escape freely.

\subsection{Astrophysical radiative shock}\label{sec:shockG}

\begin{figure}[!ht]
\centering
\includegraphics[scale=0.5]{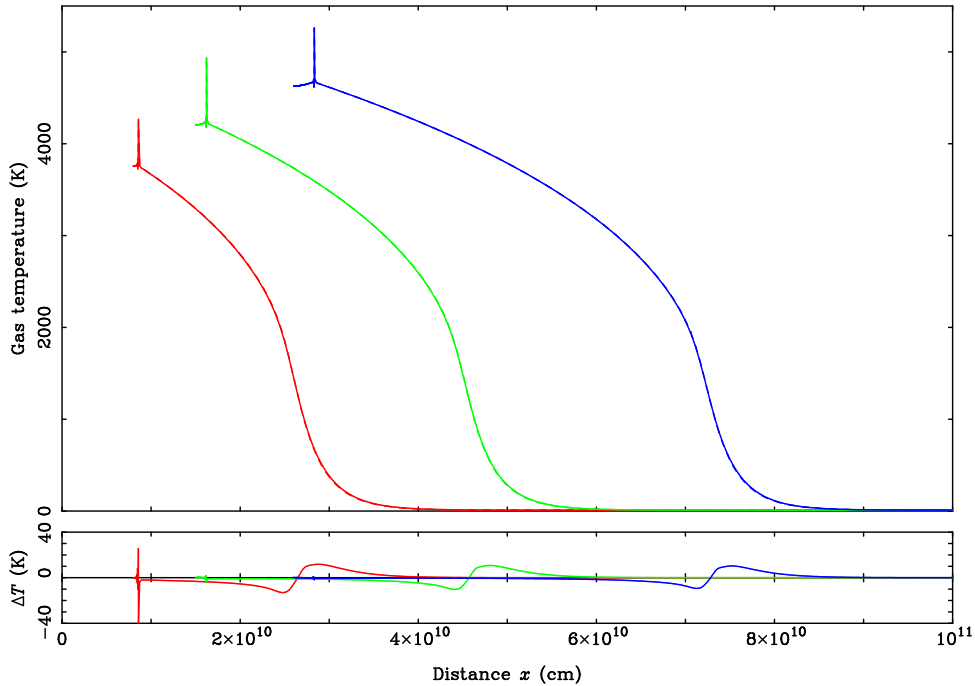}
\caption{Gas temperature in the radiative  shock test as a function of
distance at  times $t = 4.0\times10^{3}$ s  (red), $7.5\times10^{3}$ s
(green) and $1.3\times10^{4}$ s  (blue) for the monogroup run (dashed)
and  the  multigroup model  using  6  groups  (solid). The  difference
between the mono- and multigroup curves is shown in the bottom panel.}
\label{fig:shockG}
\end{figure}

The next test  is to make sure that  the multigroup radiative transfer
model is correctly  coupled to the gas hydrodynamics.  We ran a `grey'
radiative  shock simulation using  exactly the  same parameters  as in
\citet{gonzalez07}.  The  gas   inside  the  computational  domain  is
initially   at   rest   with    a   uniform   density   of   $\rho   =
7.78\times10^{-10}~\text{g~cm}^{-3}$,  temperature  $T   =  10$  K  in
equilibrium    with   the    radiation   and    opacity    $\kappa   =
0.39~\text{cm}^{2}~\text{g}^{-1}$.   The   size    of   the   box   is
$1.0\times10^{11}$ cm. We give the gas at the left boundary a velocity
of  20 km~s$^{-1}$,  which generates  the propagation  of  a radiative
shock travelling towards the right.  We use 500 equally spaced spatial
zones and 6 frequency groups (5  groups evenly spaced between $\nu = 0
- 7\times10^{14}~\text{s}^{-1}$ and  the last group  holds frequencies
from $7\times10^{14}~\text{s}^{-1}$ to $\infty$). The results at three
different epochs are shown  in Fig.~\ref{fig:shockG} (solid lines). We
have     also     run     the     radiative    shock     test     with
\textsc{sinerghy}$\mathnormal{1}$\textsc{d} using  only a single group
and the results are shown in Fig.~\ref{fig:shockG} (dashed lines). The
temperature profiles  are virtually indistinguishable,  as illustrated
by the  difference $\Delta T$  between the grey and  multigroup curves
which is  plotted below. The largest $\Delta  T$ is $\sim 40$  K for a
peak  temperature  of  4000  K,  i.e. only  one  percent.  Some  small
differences are  visible in the  radiative precursor. This  shows that
the multigroup scheme is consistent with the grey model.

We  also  note   that  the  curves  are  identical   to  the  ones  in
\citet{gonzalez07},  which  shows  that  the implicit  code  correctly
solves the equations of RHD.

\subsection{Multigroup radiative shock in xenon gas}\label{sec:shockXe}

Our final test  is to link \textsc{sinerghy}$\mathnormal{1}$\textsc{d}
to  the \textsc{odalisc}\footnote{http://irfu.cea.fr/Projets/Odalisc/}
database  of  gas  opacities  in  order  to  realistically  model  the
evolution     of     a      radiative     shock     in     a     xenon
$^{131}_{\phantom{1}54}\text{Xe}$  gas. The  \textsc{odalisc} database
aims  to  provide  spectral   opacities  as  well  as  mean  opacities
(Rosseland and Planck)  of many elements for a  wide range of physical
conditions.

The gas inside the box is  initially at rest with a uniform density of
$\rho  =   10^{-3}~\text{g~cm}^{-3}$,  temperature  $T  =   1$  eV  in
equilibrium with the radiation. The size  of the box is 36 cm with 550
zones; the first 100  cells have logarithmically increasing sizes, the
first  500 zones  cover the  range $0  - 2$  cm and  the last  50 have
steadily increasing sizes, covering the range $2 - 36$ cm. We give the
gas at the left boundary  a velocity of 60 km~s$^{-1}$, which generats
a radiative  shock travelling towards the  right. We use  an ideal gas
equation of state with atomic mass number 131.

\begin{figure}[!ht]
\centering
\includegraphics[scale=0.5]{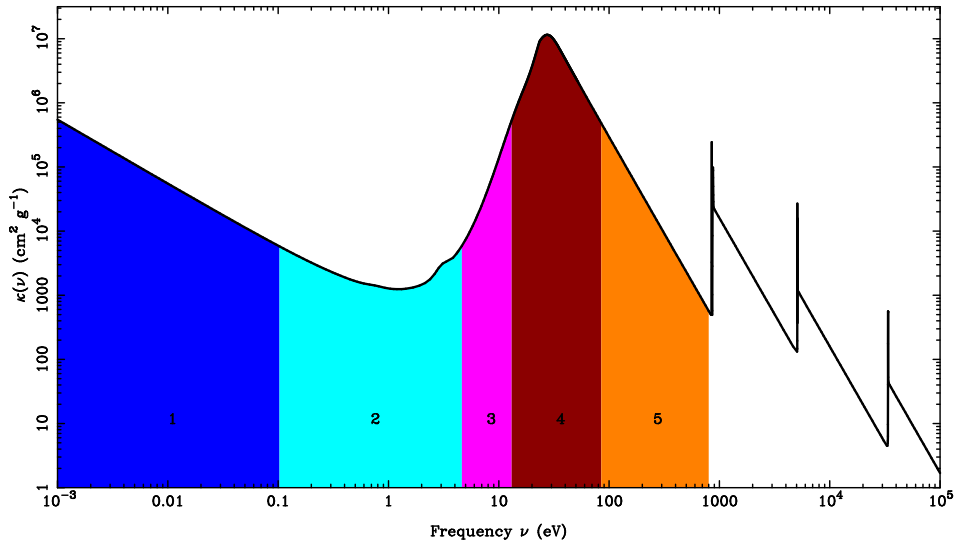}
\caption{Xenon opacities at $\rho = 10^{-3}~\text{g~cm}^{-3}$ and $T =
1$  eV  as a  function  of  frequencies.  The colours  illustrate  the
decomposition of  the frequency  domain into five  groups from  $\nu =
10^{-3}$ to 770 eV.}
\label{fig:knu}
\end{figure}

The  opacities for  the Xe  gas were  taken from  the \textsc{odalisc}
database (\textsc{gomme}  average atom model). They depend  on the gas
temperature and  density (often more strongly on  temperature) as well
as on  the frequency. The opacity  $\kappa(\nu)$ for the Xe  gas for a
density $\rho  = 10^{-3}~\text{g~cm}^{-3}$ and temperature $T  = 1$ eV
is  shown in  Fig.~\ref{fig:knu}  (black solid  curve).  We used  five
groups to  sample the opacities  from $\nu =  10^{-3}$ to 770  eV; the
colour bands in  Fig.~\ref{fig:knu} illustrate the group decomposition
of the frequency domain. Frequencies  below $10^{-3}$ eV and above 770
eV are  ignored, as  gas temperatures  in the box  remain under  30 eV
(except  the   very  narrow  temperature  spike).  The   gas  at  such
temperatures does  not radiate strongly at these  frequencies. We then
computed the Planck ($\kappa_{Pg}$) and Rosseland ($\kappa_{Rg}$) mean
opacities in each group.

Since the  gas temperature and  density evolve in time,  the opacities
need to be  calculated at each timestep in each  grid cell. The method
we used  to compute the  opacities is to  read in from the  database a
grid of  opacities for the  temperature range 0.01  to 100 eV  and the
density range $10^{-3}$ to  $0.3~\text{g~cm}^{-3}$ at the start of the
run.  From this, we  then compute  $\kappa_{Pg}$ and  $\kappa_{Rg}$ in
each  group  at  each  point  $(\rho,T)$  which  are  stored  into  an
array. During  the simulation, a  particular group opacity at  any $T$
and $\rho$ is then found using a simple four-point interpolation using
the array data.

\begin{figure}[!ht]
\centering
\includegraphics[scale=0.65]{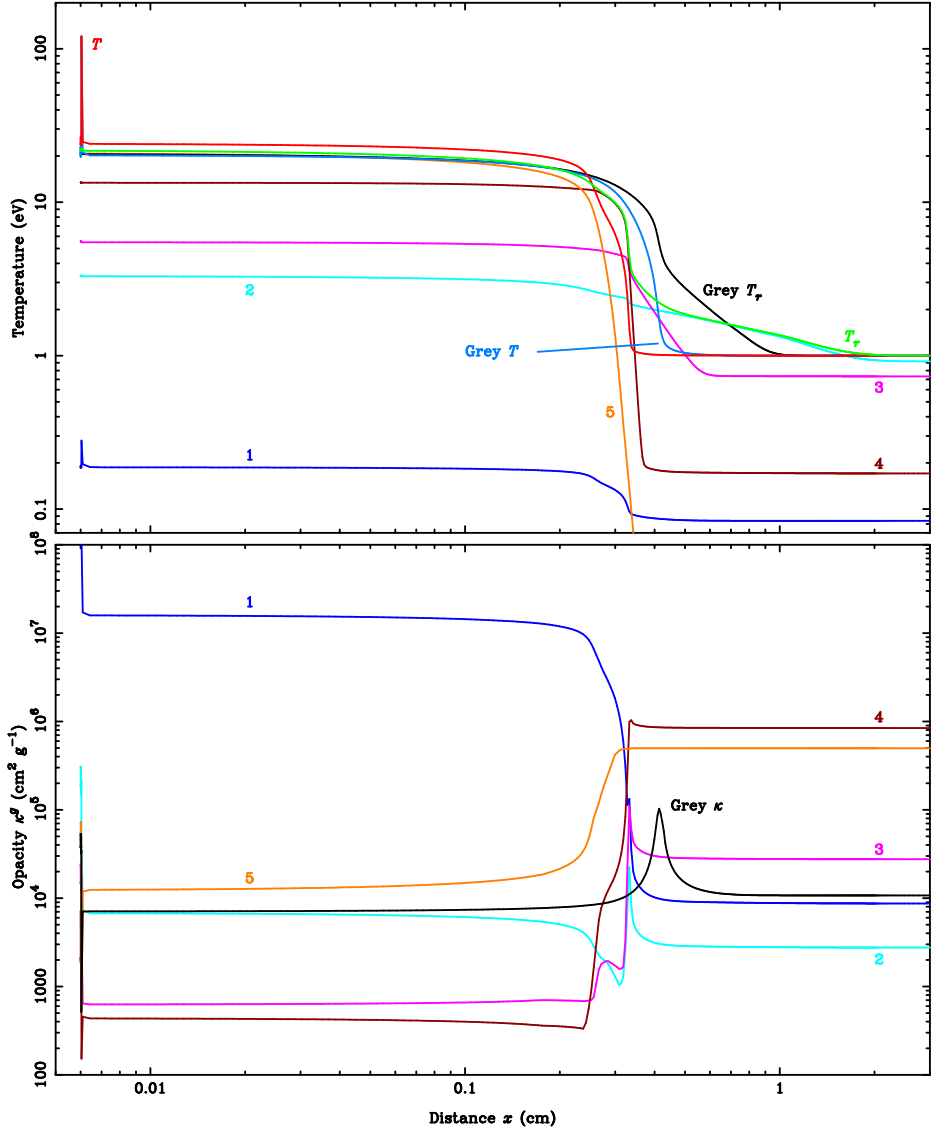}
\caption{\underline{Top  panel}:  Gas  temperature $T$  (bright  red),
total radiative  temperature $T_{r}$  (summed over all  groups, bright
green) and  individual group radiative temperatures (labelled  1 to 5)
in the multigroup simulation of a  radiative shock in Xe gas at a time
$t  = 10^{-9}$ s.  We have  also included  the gas  temperature (light
blue)    and   radiative   temperature    (black)   from    the   grey
run.  \underline{Bottom panel}:  Gas  opacities for  each  group as  a
function of  distance. The black  curve represents the opacity  in the
grey run (averaged over all frequencies).}
\label{fig:shockXe}
\end{figure}

The gas and radiative temperatures  for our simulation of a multigroup
radiative shock  in a Xe gas  at a time $t  = 10^{-9}$ s  are shown in
Fig.~\ref{fig:shockXe} (top  panel), along with  the temperatures from
an identical  but grey  run where  only a single  group over  the same
frequency range is used. A  characteristic peak in the gas temperature
(bright red) can be seen just around  $x = 6 \times 10^{-3}$ cm at the
shock.  There is  a strong  radiative precursor  (bright  green) which
extends all the way to $x  = 2$ cm. The contributions to the precursor
are clearly visible; at first  the energy from group 5 contributes the
most but subsequently  gets dominated by group 4, 3 then  2 as we move
further away from the shock.  The radiation in the low-frequency group
1 does not appear to contribute  to the dynamics of the shock. We note
that the radiation in all the  groups apart from group 2 gets absorbed
fairly  rapidly  (none  get  past  $x =0.6$  cm),  whereas  since  the
opacities  in group  2 are  the lowest  (see  Fig.~\ref{fig:knu}), the
radiation there  propagates to a greater distance.  Differences in the
positions of the  tip of the radiative precursors  in the other groups
further illustrate the effect of variable gas opacities.

Let  us note  major differences  between  the mutigroup  and the  grey
models. Due  to the fact that  opacities are averaged  over the entire
frequency range (the  low opacities are biased towards  a higher value
and  vice  versa), the  radiation  in  the  grey run  suffers  greater
absorption far away from the  shock and its radiative precursor (black
curve)  does not extend  as far  as in  the multigroup  case. However,
between 0.3 and 0.7 cm,  the grey radiative temperature is much higher
than  the multigroup  one. For  this reason,  the gas  is heated  to a
greater extent  and the  grey gas temperature  (light blue)  is higher
than the  multigroup one  around $0.3-0.4$ cm.  We also note  that the
multigroup gas temperature in the range $6 \times 10^{-3} - 0.2$ cm is
higher than for the grey run.

The opacities  of the gas in each  group and the opacity  for the grey
run are  plotted as a  function of distance  in Fig.~\ref{fig:shockXe}
(bottom panel). This is an excellent illustration of how the opacities
are affected  by the  gas temperature. For  instance, we see  that the
opacity    in    the   first    group    is    of    the   order    of
$10^{4}~\text{cm}^{2}~\text{g}^{-1}$  in  the cold  gas  ahead of  the
radiative  precursor (right  hand side)  whereas it  gains  over three
orders of magnitude in the hot  post-shock gas. For groups 3, 4 and 5,
the  opposite occurs; the  opacity is  high before  the shock  and low
after. The opacities in  the temperature transition region has diverse
behaviours. Most  importantly, the curves are very  different from the
grey opacity (black  curve) which is very constant  with a single peak
around 0.4 cm corresponding to  the jump in gas temperature (see light
blue curve in  the top panel). Interestingly, the  pre- and post-shock
grey opacities are  very similar. It becomes very  clear that the grey
model cannot  correctly represent the varied spectrum  of opacities in
such a situation,  which are crucial to the  evolution and dynamics of
the shock.

\section{Conclusions}\label{sec:con}

We have developed a multigroup  model for RHD using the $M_{1}$ moment
model. The equations of radiative  transfer are solved in the comoving
frame. In order to account for the opacity variations as a function of
frequency,  we introduced  frequency  groups and  applied the  $M_{1}$
closure inside each of them. This  gave rise to new terms depending on
the  frequency when  coupled to  the  hydrodynamics. We  use a  finite
volume method in  the frequency domain in order  to evaluate these new
coupling terms which account  for energy exchange between neighbouring
groups due  to the Doppler  effect when strong velocity  gradients are
present in the gas flow.

We  have  verified  our  method  using  a  series  of tests  for  both
radiative   transfer   alone  and   radiative   transfer  coupled   to
hydrodynamics. In the case of the radiative transfer tests, the method
was found to be successful  in reproducing the results obtained with a
kinetic code, at  a much lower computational cost.  We have shown that
the model reverts to a  grey model for frequency independent opacities
and  that the  model  is capable  of  treating the  effects of  strong
velocity gradients in a gas with frequency dependent opacities.

Finally,            we           have            coupled           the
\textsc{sinerghy}$\mathnormal{1}$\textsc{d} code to the opacities from
the   \textsc{odalisc}   database   to  realistically   simulate   the
propagation  of  a  radiative  shock  in  a Xe  gas.  We  noted  major
differences between  the multigroup and  the grey models,  showing the
importance  of accounting  for the  frequency variability  of  the gas
opacities. The  next step  in this  study will be  to use  a realistic
equation of state  for the Xe gas using  the \textsc{odalisc} database
for more realistic simulations.  \textsc{odalisc} also has a number of
different methods to calculate the opacities for each element. We will
study  the influence  of the  uncertainties  of the  opacities on  the
results  of  simulations of  radiative  shock  in  Xe in  a  following
paper.  In this work  will also  be included  an investigation  of the
impact of the  choice of frequency group boundaries  and the number of
groups on the results.

We have  now begun  the implementation of  this multigroup  method for
radiative  transfer  in  the  3D radiation  magnetohydrodynamics  code
\textsc{heracles} \citep{gonzalez07}.  The development of  such a tool
for  hydrodynamical  simulations will  prove  extremely important  for
future  studies, in  particular, in  the field  of  astrophysics where
high-energy radiation from bright stars or supernovae gets absorbed by
dense clouds  which re-radiate the  energy in the infrared,  or inside
dense stellar atmospheres where many chemical elements are present.

\section*{Acknowledgements}\label{sec:acknow}

The    authors    greatfully    acknowledge   support    from    grant
ANR-06-CIS6-009-01 for the programme SiNeRGHy. They would also like to
thank  Vladimir  Tikhonchuk   and  Matthias  Gonz\'{a}lez  for  useful
comments during the writing of this paper. We also thank the referees
for their valuable comments which have helped us improve this paper.



\end{document}